\documentclass[aps,twocolumn]{revtex4}
\usepackage{epsfig,graphics,amssymb,amsmath}
\def\dd{{\rm d}}
\def\e{{\bf e}}\def\x{{\bf x}}
\def\Pe{{\rm Pe}}

\def\xii{{{\boldsymbol \xi}}}

\begin{document}

\title{Effective diffusion of unsteady and shape-changing swimmers}
\title{Diffusing by non-swimming: Enhanced diffusion by reciprocal actuation}
\title{Enhanced diffusion by reciprocal swimming}

\author{Eric Lauga}
\email{elauga@ucsd.edu}
\affiliation{Department of Mechanical and Aerospace Engineering, University of California San Diego, 9500 Gilman Dr., La Jolla, CA 92093-0411, USA}
\date{\today}
\begin{abstract}
Purcell's scallop theorem states that swimmers deforming their shapes in a time-reversible manner (``reciprocal'' motion) cannot swim. Using numerical simulations and theoretical calculations  we show here that in a fluctuating environment,   reciprocal  swimmers undergo, on time scales larger than that of their rotational diffusion, diffusive dynamics with enhanced diffusivities, possibly by orders of magnitude, above normal translational diffusion. 
Reciprocal actuation does therefore lead to a significant advantage over non-motile behavior for small organisms such as marine bacteria.

\end{abstract}
\maketitle

In addition to its importance on our macroscopic world, fluid mechanics plays a crucial role in many cellular processes. One example is the hydrodynamics of motile cells such as bacteria, spermatozoa, algae, and half of the microorganisms on earth \cite{lighthill76,LP09}. Most of them exploit the bending or rotation of a small number of flagella (short whip-like organelles, length scale from a few to tens of microns) to create fluid-based locomotion \cite{brennen77}. In contrast, ciliated microorganisms swim by using the coordinated  beating of many short flagella termed cilia distributed along their surface \cite{brennen77}.

Two physical ideas govern the fluid mechanics of cell locomotion on small scales. The first one is the  exploitation by cells of anisotropic drag-based thrust to generate instantaneous propulsive forces \cite{LP09}. The second one is the requirement to distribute this local  propulsion along the surfaces of organisms in a manner that does not average to zero over one period of cellular actuation \cite{purcell77}. Indeed, on very small scales,  the inertia-less equations governing the surrounding fluid are linear and independent of time (Stokes equation), and thus any actuation on the fluid remaining identical under a reversal of time (so-called ``reciprocal'' actuation) cannot generate any net motion. This is known as Purcell's scallop theorem \cite{purcell77,lauga_scallop_review}.

To overcome the constraints of the scallop theorem, microorgansims swim using wave-like deformations of their appendages or bodies, be it  prokaryotes, eukaryotes  with small number of flagella, or ciliates \cite{lighthill76,LP09}. For  deformation of synthetic swimmers, at least two degrees of freedom of shape change are required  \cite{purcell77,najafi04,avron08,jellyfish}, or further 
physical effects need to be exploited, for example those leading to nonlocality (hydrodynamic interactions \cite{trouilloud08}), relaxation (actuation of flexible filaments \cite{WigginsGoldstein}) or nonlinearity (in particular, non-Newtonian stresses \cite{FuWolgemuthPowers2009}).

In contrast to large organisms able to sustain directional swimming for long periods of times, small bacteria quickly lose their orientation due to rotational Brownian motion. If $a$ is the typical hydrodynamic size of an organism in a fluid of viscosity $\eta$ and temperature $T$, this thermal orientation loss occurs on a typical time scale $\tau \sim \eta a^3 / k_B T$, of about one second for a $1~\mu$m bacterium in water, and tens of seconds for {\it E. coli}.  On time scales $t\gg \tau$,  the coupling between locomotion at a typical speed $U$ and orientation loss   \cite{lovely75,berg93} leads to  diffusive behavior for the cells with an effective diffusivity $D\sim U^2\tau$,  usually much larger than that due to normal Brownian motion.  For example, dead {\it E. coli} bacteria have diffusivities of $\approx 0.1$~$\mu$m$^2$/s while those of swimming cells are at least three orders of magnitude larger \cite{berg93}. This transition from directional motion to diffusive dynamics  was further addressed in recent work \cite{howse07}.

For small organisms significantly affected by Brownian diffusivity, we thus have  the following intriguing observation. The scallop theorem dictates how cells should deform in order to undergo  non-zero time-average displacements but at long times, cells always diffuse, and thus always display  zero time-average displacement. Would it then be possible that similar enhanced diffusive motion could be obtained within the constraints of the theorem?

In this paper we consider the fate of swimmers undergoing reciprocal actuation in a  fluctuating environment. Although the scallop theorem prevents swimming on average, we show that on time scales larger than that of rotational diffusion, these reciprocal non-swimmers undergo diffusive motion with enhanced diffusivities, possibly by orders of magnitude, above their normal Brownian diffusion in translation. This result is demonstrated  computationally using Brownian dynamics simulations, and analytically using exact  calculations for the long-time effective diffusivity  of reciprocal  unidirectional swimmers. The different regimes obtained  are also captured by physical scalings.  These new results demonstrate thus that reciprocal actuation, useless at zero temperature, does in fact lead to a significant advantage over non-motile behavior for small organisms such as marine  bacteria.  There is thus no rms scallop theorem.

\begin{figure*}[t]
\begin{center}
\includegraphics[width=0.99\textwidth]{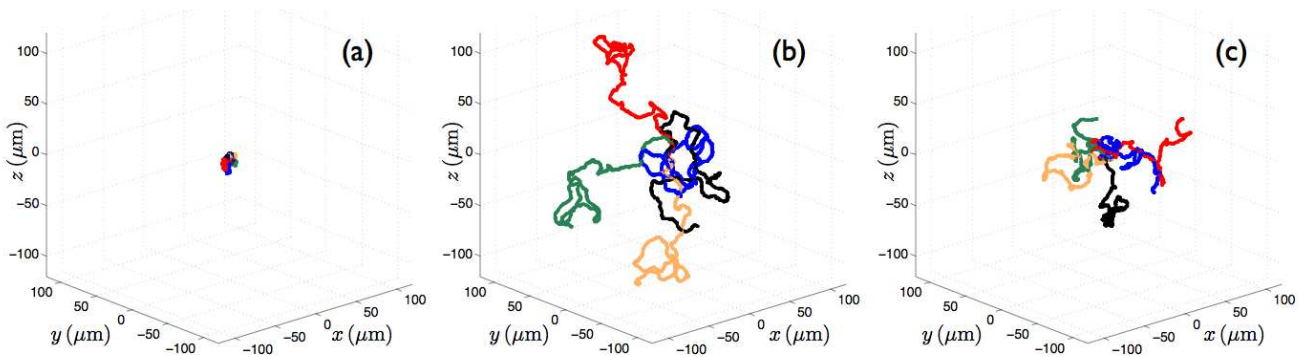}
\caption{(color online) Brownian dynamics simulation of a spherical swimmer (radius $a=1$~$\mu$m), in water at $T=300$~K during a time interval of $100$~s (5 realizations are superimposed). (a): No swimming; 
(b): Steady swimming at speed  $U=5$~$\mu$m/s;
(c) Reciprocal swimming at speed $\bar U \cos \omega t$ with $\bar U= 5$~$\mu$m/s and $\omega=2D_R$ ($D_R$ is the rotational diffusivity of the swimmer, $\omega=0.33$~rad/s).
Case (a) is pure Brownian motion while both (b) and (c) show enhanced diffusivities.}
\label{fig:BD}
\end{center}
\end{figure*}

For a first illustration of our results, we use numerical computations. We performed Brownian dynamics simulations \cite{hinch75} of a spherical swimmer (radius $a=1$~$\mu$m), in water at $T=300$~K and during a time interval of $100$~s, with results shown in Fig.~\ref{fig:BD}. The instantaneous velocity, $\bf U$, and rotation rate, $\bf \Omega$, of the sphere satisfy the dynamics: 
${\bf R}_{FU}\cdot ({\bf U}-{\bf U}_{\rm swim}) = {\bf F}^B$, 
${\bf R}_{L\Omega} \cdot {\bf \Omega} = {\bf L}^B$, 
where ${\bf U}_{\rm swim}$ is the swimming speed,
 ${\bf R}_{FU}=6\pi \eta a {\bf 1}$ and ${\bf R}_{L\Omega}=8\pi \eta a^3 {\bf 1}$ are the viscous resistances in translation and orientation (${\bf 1}$ is the identity tensor), and 
${\bf F}^B$ and ${\bf L}^B$ are, respectively, zero-mean Brownian forces and torques, with correlations governed by the  fluctuation-dissipation theorem, i.e.~$\langle{\bf F}^B(t) {\bf F}^{B}(t')^T \rangle =2k_B T {\bf R}_{FU}\delta(t-t')$ and 
$\langle{\bf L}^B(t) {\bf L}^B(t')^T \rangle =2k_B T {\bf R}_{L\Omega}\delta(t-t')$.

Simulations were performed for three different swimming behaviors; in each case  five realizations are superimposed in Fig.~\ref{fig:BD}. In  Fig.~\ref{fig:BD}a, the spheres do not swim (${\bf U}_{\rm swim}={\bf 0}$) and thus undergo pure Brownian motion. In Fig.~\ref{fig:BD}b, the spheres swim steadily at speed ${\bf U}_{\rm swim}=U {\bf e}$ where  ${\bf e}$ is a  unit vector fixed to the swimmers, and $U=5$~$\mu$m/s. With these parameters, the time scale for thermal orientation loss  is on the order of $\tau \approx 3$~s; we are thus  in the regime where $t\gg \tau$, and the steady swimmers show diffusive behavior with a diffusion constant significantly larger than the Brownian one from Fig.~\ref{fig:BD}a.  

Our new result  is illustrated in Fig.~\ref{fig:BD}c, where we show the dynamics of  swimmers undergoing reciprocal motion with  velocity ${\bf U}_{\rm swim}=U(t) {\bf e}$ and  $U(t)=\bar U \cos \omega t$ with $\bar U= 5$~$\mu$m/s and $\omega=2D_R$  where $D_R$ is the rotational diffusivity of the swimmer ($\omega=\tau^{-1}=0.33$~rad/s). Although the swimmers display no net motion even at short times  (by construction  the swimming speed averages to zero over one period of actuation), it is apparent from the numerical results that they  diffuse much faster than pure Brownian motion (Fig.~\ref{fig:BD}a). In what follows, we use scaling arguments and theoretical calculations to rationalize and quantify these results.

How can we physically account for  the increase in swimmer diffusion? The simplest approach  involves recalling the dynamics of three-dimensional (3D) random walks \cite{chandrasekhar43,doi_edwards}. If a particle at position $x$ undergoes a 3D random walk where steps of size $\ell$ are followed along random direction during  time intervals  $\delta t$, then the particle shows no average motion, $\langle x \rangle=0$, but undergoes rms spread as $\langle x^2 \rangle \sim N \ell^2$. Since time increases as  $t\sim N \delta t$, we get diffusive motion with $\langle x^2 \rangle\sim Dt$ with the diffusion constant, $D$, scaling as $D\sim \ell^2/\delta t$. In the previously-understood case of steady swimming at speed $U$, the step size is the swimming speed times the time step, $\ell = U \delta t$, and  the relevant time step for change of direction  is the time scale over which the swimming direction is lost, i.e.~$\delta t=\tau$, leading to the well-known scaling $D\sim U^2 \tau$  \cite{lovely75}.

Reciprocal non-swimmers subject to Brownian noise also behave as 3D random walkers, and to estimate  their effective diffusivity, we have to consider the appearance of a new time scale, namely the period $\omega^{-1}$ of reciprocal actuation 
over which the reversal of swimming direction occurs. We  denote by $\bar U$ the amplitude of the swimming velocity. If the period of actuation is much larger than the loss-of-orientation scale,  i.e.~$\omega^{-1}\gg \tau$, then the step size is expected to be limited by the orientation loss and scales as $\ell \sim \bar U \tau$, leading to  diffusive motion with an expected scaling $D\sim \bar U^2 \tau$. In this low-frequency limit, the effective diffusion should thus show the same scaling as the one for steady swimmers with the  velocity amplitude replacing the  steady swimming speed. In contrast, in the limit where the time for reorientation is long compared to the period of actuation, $\omega^{-1}\ll \tau$,  then the size of the 3D random walk step should be   limited by the swimming amplitude, $\ell \sim \bar U / \omega$ while the relevant time scale for change of orientation remains $\tau$, leading to an expected  high-frequency scaling for the diffusivity as $D\sim \bar U^2/\omega^2\tau$.

We now proceed to calculate exactly the effective diffusion constant for reciprocal non-swimmers in a noisy environment. We consider  instantaneous unidirectional motion with speed $U(t)$ along a direction quantified by a unit vector $\e(t)$ attached to the swimming frame -- this direction is allowed to change due to rotational diffusion. As the swimmer is subject to noise,  its position, denoted  $\x(t)$, follows, in the absence of inertia,  the dynamics 
\begin{equation}
\label{BD}
\dot\x(t) = U(t) \e(t) + \xii(t),
\end{equation}
where the zero-mean noise term $\xii$ has a magnitude set by the fluctuation dissipation theorem $\langle \xii (t)\cdot \xii (t')\rangle = 6 D_{k_BT} \delta (t-t')$. 
Here $D_{k_B T}$ is the Brownian diffusivity of the non-swimming particle ($D_{k_B T}=k_BT/6\pi\eta a$ for a sphere of radius $a$).   
In the absence of swimming ($U=0$), the swimmer displays purely Brownian motion and $\langle \x \cdot \x \rangle \approx  6 D_{k_BT} t $ in the limit $t\to \infty$. When $U\neq 0$, the swimmer position, Eq.~\eqref{BD},  can be integrated in time to give
\begin{equation}\label{int}
\x(t) = \int_0^tU(t') \e(t')\dd t' + \int_0^t \xii(t')\dd t'.
\end{equation}
The swimming direction, $\e$, varies in time according to 3D rotational diffusion \cite{doi_edwards}. We thus expect no mean direction, $\langle \e \rangle ={\bf 0}$, and an exponential loss of swimming direction over time as quantified by the correlation
\begin{equation}\label{expon}
\langle \e(t_1+t_2) \cdot \e(t_1) \rangle = e^{-t_2/\tau},
\end{equation}
with $\tau^{-1} = 2D_R $ and $D_R$ is the rotational diffusion coefficient for the swimmer ($D_R=k_B T/8\pi\eta a^3$ for a sphere). From Eq.~\eqref{BD} we thus first get that $\langle \x \rangle= {\bf 0}$ and as expected, in the long-time limit, there is no net swimming.

To quantify the effective diffusivity, we need to compute the mean square displacements. As  ${t\to\infty}$, we expect  $\langle \x \cdot \x \rangle  \approx 6 D t$ in 3D, and the effective diffusion constant, $D$,  can thus be inferred from the limit
\begin{equation}\label{Deff}
D
=\frac{1}{3}\lim_{t\to\infty} \langle \x \cdot \dot\x   \rangle.
\end{equation}
Given the integration for $\x$, Eq.~\eqref{int}, we can compute
\begin{eqnarray}
(\x\cdot \dot\x )  (t)
&=&U(t)  \left[\int_0^t\left[U(t') \e(t)\cdot\e(t') + \e(t)\cdot\xii(t')\right]\dd t'\right]\quad \nonumber\\
&& +\int_0^tU(t') \xii(t)\cdot\e(t')\dd t' + \int_0^t \xii(t)\cdot\xii(t')\dd t'.
\end{eqnarray}
Since for any times $t_1$ and $t_2$ we have no correlation
$\langle \e(t_1) \cdot \xii(t_2)\rangle = 0$, we obtain
\begin{eqnarray}
\langle \x(t) \cdot \dot\x (t)  \rangle 
=U(t)  \int_0^tU(t') \langle \e(t)\cdot\e(t')\rangle \dd t' 
+ 3 D_{k_BT},
\end{eqnarray}
which, using Eq.~\eqref{expon}, and recalling  Eq.~\eqref{Deff} leads to  
\begin{equation}\label{D}
D
=D_{k_BT}+\frac{1}{3}\left[ \lim_{t\to\infty}\int_0^t U(t) U(t') e^{-(t-t')/\tau} \dd t' \right].
\end{equation}
The effective swimmer diffusivity, Eq.~\eqref{D}, is thus given by the swimming velocity correlation function modulated by an exponential loss (for periodic swimming,  Eq.~\ref{D} should be understood as mean value over a period) \footnote{The enhanced diffusion is valid in two (2D) or three dimensions (3D). In 2D, the 1/3 coefficient in Eq.~\eqref{D} becomes 1/2, and $\tau$ is now $\tau = 1/D_R$.}.

With our exact calculation, we can now compute the effective diffusivity for some simple cases. For steady swimming $U(t)=U$, Eq.~\eqref{D} leads to 
\begin{equation}\label{D_swim}
D
=D_{k_BT}+ \frac{1}{3} U^2\tau,
\end{equation}
which is the classical result \cite{lovely75,berg93}. 
In the case of harmonic reciprocal swimming, $U(t)=\bar U\cos\omega t$, we get
\begin{equation}\label{D_reciprocal}
D=D_{k_BT}+ \frac{1}{6}\frac{\bar U^2\tau}{1+\omega^2\tau^2}\cdot
\end{equation}
More generally, for periodic swimming  of the form
$U= U_0\Re\left\{\sum_{n\geq0} a_n \exp (in\omega t)\right\}$, 
where $a_0$ is  real, we obtain 
\begin{equation}\label{final}
D
=D_{k_BT}+\frac{U_0^2 \tau}{3} 
 \left(a_0^2 + \frac{1}{2}\sum_{n\geq 1}  \frac{|a_n |^2 }{1+(n\tau\omega)^2} 
 \right),
\end{equation}
which clearly displays both scalings  for $\omega \tau \gg 1$ and $\omega \tau \ll 1$ discussed above. We also get from 
Eq.~\eqref{final} that we have always have $D > D_{k_BT}$. For example, for a periodic square swimming  with
$U(t)=-\bar U$ during $t \in (-\pi/\omega,0)$ and then instantaneous reversal $U(t)=+\bar U$ for  $t\in (0,\pi/\omega)$, we have $U_0=\bar U$, 
$a_{2p}=0$ and $a_{2p+1}= {-4i}/{\pi(2p+1)}$,
leading to
\begin{equation}
D=D_{k_BT}
  +  \frac{ \bar U ^2  \tau}{3}  
{ \left[1 -\frac{2 \tau\omega}{\pi} \tanh\left(\frac{\pi}{ 2 \tau\omega}\right)\right]}\cdot
 \end{equation}

\begin{figure}[t]
\begin{center}
\includegraphics[width=0.48\textwidth]{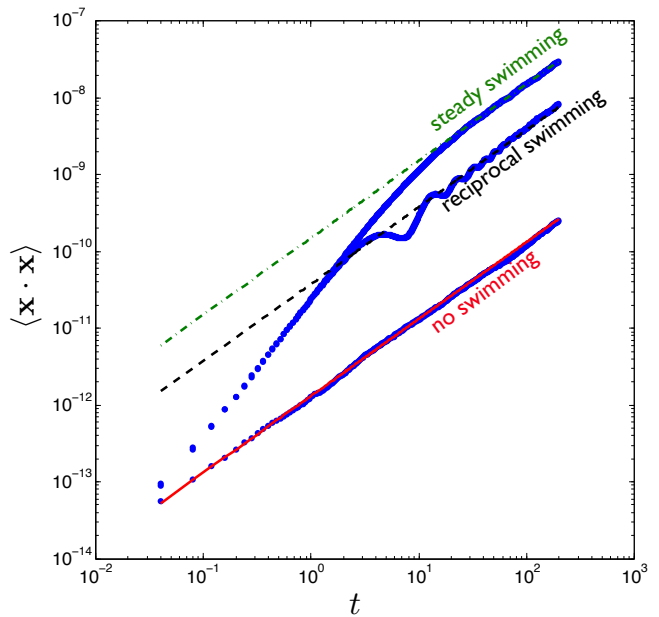}
\caption{(color online) Comparison between simulations and theoretical predictions
for a spherical swimmer.  
Symbols:  Brownian dynamics  simulations for the same three cases as in Fig.~\ref{fig:BD} (averages of 500 realizations over a time interval of 200 seconds). 
From top to bottom: steady, reciprocal and no swimming.
Theoretical predictions  shown as straight lines.  Top (green dash-dotted line): prediction for effective diffusion for steady swimming, Eq.~\eqref{D_swim};
middle (black dashed line):  prediction for diffusion by reciprocal swimming, Eq.~\eqref{D_reciprocal}; bottom (red solid line): Brownian motion.}
\label{fig:BD2}
\end{center}
\end{figure}

In  Fig.~\ref{fig:BD2} we show a comparison between our analytical predictions and our computational results. We plot the mean square displacement of 500 realizations of the swimmers with the same three cases as in Fig.~\ref{fig:BD} over a time interval of $ 200 s$.  For the three cases considered (no swimming, steady swimming, and reciprocal swimming), we also plot as straight lines the theoretical prediction (where $D$ is given, respectively, by $D_{k_B T}$, Eq.~\ref{D_swim} and Eq.~\ref{D_reciprocal}). We obtain excellent quantitative agreement, confirming the validity of our theoretical approach.

Biologically, our results are relevant to the dynamics of marine bacteria. Non-marine bacteria such as {\it E. coli}  swim using a ``run-and-tumble'' strategy where straight swimming paths are followed by random re-orientation events \cite{bergbook}. As a difference, marine bacteria display ``run-and-reverse'' (or ``back-and-forth'') locomotion where 
high speed swimming  along  straight paths is followed by almost complete reversal of their swimming direction  \cite{mitchell96,johansen02}. With no bias in the characteristics of the paths, this is the example of a biological reciprocal swimmer.

To estimate the order of magnitude of our result, let us consider an elongated bacterium  characterized by two length scales, $b$ and $a\gg b$. Scaling-wise, we have 
$D_{k_BT}\sim{k_BT}/{\eta a \log(a/b)}$, 
$D_{R}\sim {k_BT}/{\eta a^3 \log(a/b)}$,
and thus the reorientation time scales as $\tau \sim a^2 /D_{k_BT}$. 
The maximum enhanced diffusivity is obtained in the low frequency limit, $\omega\tau \lesssim 1$. In that case, the increase of cell diffusivity, in a quiescent fluid environment,  is given by 
${D}/{D_{k_BT}}\sim {\bar U^2 \tau }/{ D_{k_BT}}\sim \Pe^2$,
where the Peclet number is given by $\Pe=a\bar U/D_{k_BT}$. For blunt swimmers where $a \approx b$, even though  the log terms in  the diffusion constants  disappear, the result  is  unchanged. For order one or above Peclet numbers, the diffusive behavior of cells is thus expected to be dominated by all swimming-induced terms, including the reciprocal ones. For a ten micron  bacterium in water at room temprature, this corresponds to a critical amplitude of reciprocal swimming of  $\bar U  \approx 10\,$nm/s, less than 0.1\% of the steady swimming speed of most marine bacteria  \cite{johansen02}. 
For example, the micron-size marine bacterium {\it Shewanella putrefaciens} (CN32) has an average swimming speed of 100 $\mu$m/s and run duration of about 1 second \cite{mitchell96}, leading to an expected reciprocal diffusivity of 10 $\mu$m$^2$/s, over two orders of magnitude above that given by Brownian motion.

Many marine bacteria are found in high-Reynolds number  turbulent fluid environments \cite{luchsinger99}. The framework used here  remains valid provided $T$ is interpreted as an effective temperature, with an equation equivalent to Eq.~\eqref{expon} capturing the rotational dynamics of bacteria in turbulent flows. The results reported  in this paper could thus be used to describe the effective diffusion of marine bacteria in intermittent or turbulent  flows. Our work could also be adapted to describe biased effective diffusion and chemotaxis in presence of external fields, for example if we allow the reciprocal swimming amplitude, or its frequency, to be coupled to an external chemical concentration. More generally, any noisy process leading to an exponential loss of cell orientation will lead to enhanced diffusion for reciprocal actuation, for example cell-cell collisions at high density \cite{ishikawa_pedley_diffusion07}.

In summary, we have shown in this paper that reciprocal swimmers, previously believed to display a useless form of locomotion,  undergo in fact enhanced diffusion, possibly by orders of magnitude, over inert bodies of the same size. Purcell's scallop theorem, valid in the absence of noise, can therefore not be extended in a fluctuating environment, and reciprocal (or more generally, unsteady) actuation can lead to significant advantages over non-motile behavior for small organisms.

\vspace{0.5mm}
We thank Tim Pedley for helpful comments. Funding by the National Science Foundation (grant CBET-0746285) is gratefully acknowledged.

\bibliography{bib_2010}

\begin{thebibliography}{34}
\expandafter\ifx\csname natexlab\endcsname\relax\def\natexlab#1{#1}\fi
\expandafter\ifx\csname bibnamefont\endcsname\relax
  \def\bibnamefont#1{#1}\fi
\expandafter\ifx\csname bibfnamefont\endcsname\relax
  \def\bibfnamefont#1{#1}\fi
\expandafter\ifx\csname citenamefont\endcsname\relax
  \def\citenamefont#1{#1}\fi
\expandafter\ifx\csname url\endcsname\relax
  \def\url#1{\texttt{#1}}\fi
\expandafter\ifx\csname urlprefix\endcsname\relax\def\urlprefix{URL }\fi
\providecommand{\bibinfo}[2]{#2}
\providecommand{\eprint}[2][]{\url{#2}}

\bibitem[{\citenamefont{Lighthill}(1976)}]{lighthill76}
\bibinfo{author}{\bibfnamefont{J.}~\bibnamefont{Lighthill}},
  \bibinfo{journal}{SIAM Rev.} \textbf{\bibinfo{volume}{18}},
  \bibinfo{pages}{161} (\bibinfo{year}{1976}).

\bibitem[{\citenamefont{Lauga and Powers}(2009)}]{LP09}
\bibinfo{author}{\bibfnamefont{E.}~\bibnamefont{Lauga}} \bibnamefont{and}
  \bibinfo{author}{\bibfnamefont{T.~R.} \bibnamefont{Powers}},
  \bibinfo{journal}{Rep. Prog. Phys.} \textbf{\bibinfo{volume}{72}},
  \bibinfo{pages}{096601} (\bibinfo{year}{2009}).

\bibitem[{\citenamefont{Brennen and Winet}(1977)}]{brennen77}
\bibinfo{author}{\bibfnamefont{C.}~\bibnamefont{Brennen}} \bibnamefont{and}
  \bibinfo{author}{\bibfnamefont{H.}~\bibnamefont{Winet}},
  \bibinfo{journal}{Ann. Rev. Fluid Mech.} \textbf{\bibinfo{volume}{9}},
  \bibinfo{pages}{339} (\bibinfo{year}{1977}).

\bibitem[{\citenamefont{Purcell}(1977)}]{purcell77}
\bibinfo{author}{\bibfnamefont{E.~M.} \bibnamefont{Purcell}},
  \bibinfo{journal}{Am. J. Phys.} \textbf{\bibinfo{volume}{45}},
  \bibinfo{pages}{3} (\bibinfo{year}{1977}).

\bibitem[{\citenamefont{Lauga}(2011)}]{lauga_scallop_review}
\bibinfo{author}{\bibfnamefont{E.}~\bibnamefont{Lauga}}, \bibinfo{journal}{Soft
  Matt. - DOI: 10.1039/c0sm00953a}  (\bibinfo{year}{2011}).

\bibitem[{\citenamefont{Najafi and Golestanian}(2004)}]{najafi04}
\bibinfo{author}{\bibfnamefont{A.}~\bibnamefont{Najafi}} \bibnamefont{and}
  \bibinfo{author}{\bibfnamefont{R.}~\bibnamefont{Golestanian}},
  \bibinfo{journal}{Phys. Rev. E} \textbf{\bibinfo{volume}{69}},
  \bibinfo{pages}{062901} (\bibinfo{year}{2004}).

\bibitem[{\citenamefont{Avron and Raz}(2008)}]{avron08}
\bibinfo{author}{\bibfnamefont{J.~E.} \bibnamefont{Avron}} \bibnamefont{and}
  \bibinfo{author}{\bibfnamefont{O.}~\bibnamefont{Raz}}, \bibinfo{journal}{New
  J. Phys.} \textbf{\bibinfo{volume}{10}}, \bibinfo{pages}{063016}
  (\bibinfo{year}{2008}).

\bibitem[{\citenamefont{Evans et~al.}(2010)\citenamefont{Evans, Spagnolie, and
  Lauga}}]{jellyfish}
\bibinfo{author}{\bibfnamefont{A.~A.} \bibnamefont{Evans}},
  \bibinfo{author}{\bibfnamefont{S.~E.} \bibnamefont{Spagnolie}},
  \bibnamefont{and} \bibinfo{author}{\bibfnamefont{E.}~\bibnamefont{Lauga}},
  \bibinfo{journal}{Soft Matt.} \textbf{\bibinfo{volume}{6}},
  \bibinfo{pages}{1737} (\bibinfo{year}{2010}).

\bibitem[{\citenamefont{Trouilloud et~al.}(2008)\citenamefont{Trouilloud, Yu,
  Hosoi, and Lauga}}]{trouilloud08}
\bibinfo{author}{\bibfnamefont{R.}~\bibnamefont{Trouilloud}},
  \bibinfo{author}{\bibfnamefont{T.~S.} \bibnamefont{Yu}},
  \bibinfo{author}{\bibfnamefont{A.~E.} \bibnamefont{Hosoi}}, \bibnamefont{and}
  \bibinfo{author}{\bibfnamefont{E.}~\bibnamefont{Lauga}},
  \bibinfo{journal}{Phys. Rev. Lett.} \textbf{\bibinfo{volume}{101}},
  \bibinfo{pages}{048102} (\bibinfo{year}{2008}); 
\bibinfo{author}{\bibfnamefont{E.}~\bibnamefont{Lauga}} \bibnamefont{and}
  \bibinfo{author}{\bibfnamefont{D.}~\bibnamefont{Bartolo}},
  \bibinfo{journal}{Phys. Rev. E} \textbf{\bibinfo{volume}{78}},
  \bibinfo{pages}{030901} (\bibinfo{year}{2008});
\bibinfo{author}{\bibfnamefont{G.~P.} \bibnamefont{Alexander}}
  \bibnamefont{and} \bibinfo{author}{\bibfnamefont{J.~M.}
  \bibnamefont{Yeomans}}, \bibinfo{journal}{Euro. Phys. Lett.}
  \textbf{\bibinfo{volume}{83}}, \bibinfo{pages}{34006} (\bibinfo{year}{2008}).

\bibitem[{\citenamefont{Wiggins and Goldstein}(1998)}]{WigginsGoldstein}
\bibinfo{author}{\bibfnamefont{C.~H.} \bibnamefont{Wiggins}} \bibnamefont{and}
  \bibinfo{author}{\bibfnamefont{R.~E.} \bibnamefont{Goldstein}},
  \bibinfo{journal}{Phys. Rev. Lett.} \textbf{\bibinfo{volume}{80}},
  \bibinfo{pages}{3879 } (\bibinfo{year}{1998});
\bibinfo{author}{\bibfnamefont{E.}~\bibnamefont{Lauga}},
  \bibinfo{journal}{Phys. Rev. E} \textbf{\bibinfo{volume}{75}},
  \bibinfo{pages}{041916} (\bibinfo{year}{2007});
\bibinfo{author}{\bibfnamefont{R.}~\bibnamefont{Dreyfus}},
  \bibinfo{author}{\bibfnamefont{J.}~\bibnamefont{Baudry}},
  \bibinfo{author}{\bibfnamefont{M.~L.} \bibnamefont{Roper}},
  \bibinfo{author}{\bibfnamefont{M.}~\bibnamefont{Fermigier}},
  \bibinfo{author}{\bibfnamefont{H.~A.} \bibnamefont{Stone}}, \bibnamefont{and}
  \bibinfo{author}{\bibfnamefont{J.}~\bibnamefont{Bibette}},
  \bibinfo{journal}{Nature} \textbf{\bibinfo{volume}{437}},
  \bibinfo{pages}{862} (\bibinfo{year}{2005}).

\bibitem[{\citenamefont{Fu et~al.}(2009)\citenamefont{Fu, Wolgemuth, and  Powers}}]{FuWolgemuthPowers2009}
\bibinfo{author}{\bibfnamefont{H.~C.} \bibnamefont{Fu}},
  \bibinfo{author}{\bibfnamefont{C.~W.} \bibnamefont{Wolgemuth}},
  \bibnamefont{and} \bibinfo{author}{\bibfnamefont{T.~R.}
  \bibnamefont{Powers}}, \bibinfo{journal}{Phys. Fluids}
  \textbf{\bibinfo{volume}{21}}, \bibinfo{pages}{033102}
  (\bibinfo{year}{2009});
\bibinfo{author}{\bibfnamefont{E.}~\bibnamefont{Lauga}},
  \bibinfo{journal}{Europhys. Lett.} \textbf{\bibinfo{volume}{86}},
  \bibinfo{pages}{64001} (\bibinfo{year}{2009});
\bibinfo{author}{\bibfnamefont{O.~S.} \bibnamefont{Pak}},
  \bibinfo{author}{\bibfnamefont{T.}~\bibnamefont{Normand}}, \bibnamefont{and}
  \bibinfo{author}{\bibfnamefont{E.}~\bibnamefont{Lauga}},
  \bibinfo{journal}{Phys. Rev. E} \textbf{\bibinfo{volume}{81}},
  \bibinfo{pages}{036312} (\bibinfo{year}{2010}).

\bibitem[{\citenamefont{Lovely and Dahlquist}(1975)}]{lovely75}
\bibinfo{author}{\bibfnamefont{P.~S.} \bibnamefont{Lovely}} \bibnamefont{and}
  \bibinfo{author}{\bibfnamefont{F.~W.} \bibnamefont{Dahlquist}},
  \bibinfo{journal}{J. Theor. Biol.} \textbf{\bibinfo{volume}{50}},
  \bibinfo{pages}{477} (\bibinfo{year}{1975}).

\bibitem[{\citenamefont{Berg}(1993)}]{berg93}
\bibinfo{author}{\bibfnamefont{H.~C.} \bibnamefont{Berg}},
  \emph{\bibinfo{title}{Random walks in biology}}
  (\bibinfo{publisher}{Princeton University Press},
  \bibinfo{address}{Princeton, N. J.}, \bibinfo{year}{1993}).

\bibitem[{\citenamefont{Howse et~al.}(2007)\citenamefont{Howse, Jones, Ryan,
  Gough, Vafabakhsh, and Golestanian}}]{howse07}
\bibinfo{author}{\bibfnamefont{J.~R.} \bibnamefont{Howse}},
  \bibinfo{author}{\bibfnamefont{R.~A.~L.} \bibnamefont{Jones}},
  \bibinfo{author}{\bibfnamefont{A.~J.} \bibnamefont{Ryan}},
  \bibinfo{author}{\bibfnamefont{T.}~\bibnamefont{Gough}},
  \bibinfo{author}{\bibfnamefont{R.}~\bibnamefont{Vafabakhsh}},
  \bibnamefont{and}
  \bibinfo{author}{\bibfnamefont{R.}~\bibnamefont{Golestanian}},
  \bibinfo{journal}{Phys. Rev. Lett.} \textbf{\bibinfo{volume}{99}},
  \bibinfo{pages}{048102} (\bibinfo{year}{2007});
\bibinfo{author}{\bibfnamefont{V.}~\bibnamefont{Lobaskin}},
  \bibinfo{author}{\bibfnamefont{D.}~\bibnamefont{Lobaskin}}, \bibnamefont{and}
  \bibinfo{author}{\bibfnamefont{I.~M.} \bibnamefont{Kulic}},
  \bibinfo{journal}{Eur. Phys. J. - Sp. Topics} \textbf{\bibinfo{volume}{157}},
  \bibinfo{pages}{149} (\bibinfo{year}{2008});
\bibinfo{author}{\bibfnamefont{J.}~\bibnamefont{Dunkel}} \bibnamefont{and}
  \bibinfo{author}{\bibfnamefont{I.~M.} \bibnamefont{Zaid}},
  \bibinfo{journal}{Phys. Rev. E} \textbf{\bibinfo{volume}{80}},
  \bibinfo{pages}{021903} (\bibinfo{year}{2009});
\bibinfo{author}{\bibfnamefont{B.}~\bibnamefont{ten Hagen}},
  \bibinfo{author}{\bibfnamefont{S.}~\bibnamefont{van Teeffelen}},
  \bibnamefont{and} \bibinfo{author}{\bibfnamefont{H.}~\bibnamefont{L\"owen}},
  \bibinfo{journal}{Preprint http://arxiv.org/abs/1005.1343}
  (\bibinfo{year}{2011});
\bibinfo{author}{\bibfnamefont{M.}~\bibnamefont{Garcia}},
  \bibinfo{author}{\bibfnamefont{S.}~\bibnamefont{Berti}},
  \bibinfo{author}{\bibfnamefont{P.}~\bibnamefont{Peyla}}, \bibnamefont{and}
  \bibinfo{author}{\bibfnamefont{S.}~\bibnamefont{Rafa•}},
  \bibinfo{journal}{Preprint http://arxiv.org/abs/1011.2931}
  (\bibinfo{year}{2011}).

\bibitem[{\citenamefont{Hinch}(1975)}]{hinch75}
\bibinfo{author}{\bibfnamefont{E.~J.} \bibnamefont{Hinch}},
  \bibinfo{journal}{J. Fluid Mech.} \textbf{\bibinfo{volume}{72}},
  \bibinfo{pages}{499} (\bibinfo{year}{1975});
\bibinfo{author}{\bibfnamefont{D.~L.} \bibnamefont{Ermak}} \bibnamefont{and}
  \bibinfo{author}{\bibfnamefont{J.~A.} \bibnamefont{McCammon}},
  \bibinfo{journal}{J. Chem. Phys.} \textbf{\bibinfo{volume}{69}},
  \bibinfo{pages}{1352} (\bibinfo{year}{1978}).

\bibitem[{\citenamefont{Chandrasekhar}(1943)}]{chandrasekhar43}
\bibinfo{author}{\bibfnamefont{S.}~\bibnamefont{Chandrasekhar}},
  \bibinfo{journal}{Rev. Mod. Phys.} \textbf{\bibinfo{volume}{15}},
  \bibinfo{pages}{1} (\bibinfo{year}{1943}); 
\bibinfo{author}{\bibfnamefont{E.~A.} \bibnamefont{Codling}},
  \bibinfo{author}{\bibfnamefont{M.~J.} \bibnamefont{Plank}}, \bibnamefont{and}
  \bibinfo{author}{\bibfnamefont{S.}~\bibnamefont{Benhamou}},
  \bibinfo{journal}{J. Roy. Soc. Int.} \textbf{\bibinfo{volume}{5}},
  \bibinfo{pages}{813} (\bibinfo{year}{2008}).

\bibitem[{\citenamefont{Doi and Edwards}(1999)}]{doi_edwards}
\bibinfo{author}{\bibfnamefont{M.}~\bibnamefont{Doi}} \bibnamefont{and}
  \bibinfo{author}{\bibfnamefont{S.~F.} \bibnamefont{Edwards}},
  \emph{\bibinfo{title}{The Theory of Polymer Dynamics}}
  (\bibinfo{publisher}{Clarendon Press}, \bibinfo{address}{Oxford, U. K.},
  \bibinfo{year}{1999}).


\bibitem[{\citenamefont{Berg}(2004)}]{bergbook}
\bibinfo{author}{\bibfnamefont{H.~C.} \bibnamefont{Berg}},
  \emph{\bibinfo{title}{{\it E. coli} in {M}otion}}
  (\bibinfo{publisher}{Springer-Verlag}, \bibinfo{address}{New York, NY},
  \bibinfo{year}{2004}).

\bibitem[{\citenamefont{Mitchell et~al.}(1996)\citenamefont{Mitchell, Pearson,
  and Dillon}}]{mitchell96}
\bibinfo{author}{\bibfnamefont{J.~G.} \bibnamefont{Mitchell}},
  \bibinfo{author}{\bibfnamefont{L.}~\bibnamefont{Pearson}}, \bibnamefont{and}
  \bibinfo{author}{\bibfnamefont{S.}~\bibnamefont{Dillon}},
  \bibinfo{journal}{Appl. Environ. Microbiol.} \textbf{\bibinfo{volume}{62}},
  \bibinfo{pages}{3716} (\bibinfo{year}{1996}).

\bibitem[{\citenamefont{Johansen et~al.}(2002)\citenamefont{Johansen, Pinhassi,
  Blackburn, Zweifel, and Hagstrom}}]{johansen02}
\bibinfo{author}{\bibfnamefont{J.~E.} \bibnamefont{Johansen}},
  \bibinfo{author}{\bibfnamefont{J.}~\bibnamefont{Pinhassi}},
  \bibinfo{author}{\bibfnamefont{N.}~\bibnamefont{Blackburn}},
  \bibinfo{author}{\bibfnamefont{U.~L.} \bibnamefont{Zweifel}},
  \bibnamefont{and} \bibinfo{author}{\bibfnamefont{A.}~\bibnamefont{Hagstrom}},
  \bibinfo{journal}{Aqu. Microb. Ecol.} \textbf{\bibinfo{volume}{28}},
  \bibinfo{pages}{229} (\bibinfo{year}{2002}).

\bibitem[{\citenamefont{Luchsinger et~al.}(1999)\citenamefont{Luchsinger,
  Bergersen, and Mitchell}}]{luchsinger99}
\bibinfo{author}{\bibfnamefont{R.~H.} \bibnamefont{Luchsinger}},
  \bibinfo{author}{\bibfnamefont{B.}~\bibnamefont{Bergersen}},
  \bibnamefont{and} \bibinfo{author}{\bibfnamefont{J.~G.}
  \bibnamefont{Mitchell}}, \bibinfo{journal}{Biophys. J.}
  \textbf{\bibinfo{volume}{77}}, \bibinfo{pages}{2377} (\bibinfo{year}{1999}).

\bibitem[{\citenamefont{Ishikawa and
  Pedley}(2007)}]{ishikawa_pedley_diffusion07}
\bibinfo{author}{\bibfnamefont{T.}~\bibnamefont{Ishikawa}} \bibnamefont{and}
  \bibinfo{author}{\bibfnamefont{T.~J.} \bibnamefont{Pedley}},
  \bibinfo{journal}{J. Fluid Mech.} \textbf{\bibinfo{volume}{588}},
  \bibinfo{pages}{437} (\bibinfo{year}{2007}).

\end{thebibliography}
\end{document}